\documentstyle[12pt]{article}
\def\a{\begin{eqnarray}}
\def\b{\end{eqnarray}}
\def\0{\nonumber}
\def\ba{\begin{array}}
\def\ea{\end{array}}
\def\al{{\alpha}}
\def\lm{{\lambda}}
\def\d{{\partial}}
\def\Tr{{\rm Tr}}

\begin{document}
\begin{flushright}
SISSA-ISAS 92/95/EP\\
hep-th/9511076
\end{flushright}
\vskip0.5cm
\centerline{\LARGE\bf Star--Matrix Models }
\vskip0.3cm
\centerline{\large   E.Vinteler}
\centerline{International School for Advanced Studies (SISSA/ISAS)}
\centerline{Via Beirut 2, 34014 Trieste, Italy}
\centerline{e-mail:vinteler@gandalf.sissa.it}
\vskip0.5cm
\vskip5cm
\abstract{The star-matrix models are difficult to solve due to the
 multiple  powers  of  the  Vandermonde  determinants  in  the
partition function. We apply  to these models a modified  Q-matrix approach
and we get results consistent with those obtained by other methods.As
examples we study the inhomogenous gaussian model on Bethe tree and
matrix $q$-Potts-like model. For the last model in the
special cases $q=2$ and $q=3$, we write down explicit formulas which
determinate the critical behaviour of the system.For $q=2$ we argue
that the critical behaviour is indeed that of the Ising model on the
$\phi^3$ lattice. }
\vfill\eject

\section{Introduction}
The    star-matrix    models    were    considered    first    in
\cite{Kaz} \cite{KM1}  in  connection  with  the  q-Potts   model
and percolation problem.Another field of interest is  the  "induced"  QCD
which in the large $N$ limit is  equivalent  with  the  star-model  on
Bethe tree. The star-models are also a direct  generalization  of  the
$c=1$ matrix model, which is the particular case for $q=1$.

 The difficulties in solving exactly these models are related
with the multiple  powers  of  the  Vandermonde  determinants  in  the
partition function.In the Q-matrix approach, another problem
 related with the first , is  the
choosing of a proper  basis  in  which  to  define  consistently  the
Q-matrices.

In this work we show that it is possible to apply a modified  Q-matrix
approach and to  define  consistently   the  Q-matrices.  The   method
gives  consistent  results for the cases where    other   methods  can
be  applied  :the  saddle-point  method,   Schwinger-Dyson   approach,
gaussian integration etc. The gaussian  models  and  matrix  model  on
Bethe tree give similar results with the previous ones.  However,  our
method can be applied  in all genera, and permits a more precise study
of particular parts of the Bethe tree.

 In the last section  we
consider the matrix formulation of the q-Potts-like model and study
it as a polymer on a  random
surface. It is important that we  still get  consistent  results  for  this
model. It is a non-trivial result,  because  the  coupling  conditions
are overdetermined: the Q-matrices depend on $3q+3$ variables  and  we
have $4q+3$ equations. More generally  when  the   central   potential
$V_0$ is
gaussian and the lateral potentials $V_\al$ are of order $n$ we have  $3q+n$
variables  and  $(n+1)q+3$ equations always with combatible solutions.

\section{Q-matrix approach}
\setcounter{equation}{0}
\setcounter{subsection}{0}
The Q-matrix approach was introduced for the 1-matrix model \cite{IM}
and
further developed for the two-matrix  and the chain multi-matrix models in
\cite{BX}.
Here we are applying this approach to another class of multi-matrix
models -- the star-matrix models.

The partition function of the $q$-star model is given by:
\a
Z=\int \prod_{\al=1}^q dM_\al dM_0 \exp(V_0+\sum_{\al}^q V_\al+
M_0\sum_{\al=1}^q c_\al M_\al)
\b
with  the potentials $V_\al=\sum_{r=1}^{p_\al}t_{\al,r}M_\al^r,\al=0,1\ldots
q$.
It is possible to integrate over the angular degrees of freedom
and to remain with integrals only over the eigenvalues:
\a
Z=\int \prod_{i=1}^N(d\lm_{0,i}\prod_{\al=1}^q d\lm_{\al,i})(\Delta
(\lm_0))^{2-q}\prod_{\al=1}^q
\Delta(\lm_\al)e^V
\b
with
\a
V=\sum_{i=1}^N( V_0(\lm_{0,i})+\sum_{\al=1}^q V_\al(\lm_{\al,i})+
\lm_{0,i}\sum_{\al=1}^q c_\al\lm_{\al,i}) \0
\b
We define the orthogonal functions basis as $\xi_n$ and $q+1$-conjugate
functions $\eta_{\al,m}$:
\a
\int d\lm_0\prod_{\al=1}^q d\lm_{\al}\xi_n^q(\lm_0)e^V
\prod_{\al=1}^q\eta_{\al,m_\al}(\lm_\al)=h_{n}\delta_{nm}
\prod_{\al=1}^q\delta_{m,m_\al}
\label{ort}
\b
The basic functions have the property:
\a
{h_{o,m}\over h_{\al,m}}\eta_{0,m}(\lm_0)=
\int d\lm_\al e^{V_\al+c_\al\lm_0\lm_\al}\eta_{\al,m}(\lm_\al)\label{int}
\b
Introducing this relation (\ref{int}) in the orthogonality condition
(\ref{ort}) we get:
\a
\int d\lm_0\xi_{0,n}^q(\lm_0)e^{ V_0}\prod_{\al=1}^q\eta_{0,m_\al}^q(\lm_0)=
h_{n}\prod_{\al=1}^q \delta_{n,m_\al}
\label{ort1}
\b
Because the basic functions $\xi_{0,n}(\lm_0), \eta_{0,n}(\lm_0)$ at
power $q$ are linear combinations of basic functions:
\a
\xi_{0,n}^q(\lm_0)=\xi_{0,nq}(\lm_0)+\sum_{k=1}^{nq}a_k\xi_{0,nq}(\lm_0),\
\eta_{0,n}^q(\lm_0)=\eta_{0,nq}(\lm_0)+\sum_{k=1}^{nq}
{\overline a_k}\eta_{0,nq}(\lm_0)
\b
we can show that the integral (\ref{ort1}) follows from the orthogonal
condition :
\a
\int d\lm_o \xi_{0,n}(\lm_0)e^{ V_0}\eta_{0,m}(\lm_0)=h_{n}\delta_{n,m}
\b
and:
\a
h_n=n_{0,nq}+\sum_{k=1}^{nq}|a_k|^2 h_{0,nq-k}\0
\b
Inserting the expression of $\eta_{0,m}$ (\ref{int}) in the relation
(\ref{ort1})we get  the orthogonal condition:
\a
\int d\lm_0 d\lm_\al \xi_{0,n}(\lm_0)e^{ V_0+V_\al+c_\al\lm_0\lm_\al}
\eta_{\al,m}^q(\lm_\al)=h_{\al,n}\delta_{n,m}
\label{orthal}
\b

The property (\ref{int}) gives the possibility to integrate over
Vandermonde determinants:
\a
\prod_{i=0}^{N-1}{h_{\al,i}\over
h_{0,i}}\det_{ij}[\eta_{0,i}(\lm_{0,j})]=
\int d\lm_\al e^{V_\al+c_\al\lm_0\lm_\al}\det_{ij}[\eta_{\al,i}(\lm_{\al,j})]\0
\b
and permits to calculate the partition function:
\a
Z=const N!\prod_{i=0}^{N-1}h_{0,i}^{1-q}\left(\prod_{\al=1}^q h_{\al,i}\right)
\b
We introduce the $Q$-matrices as:

\a
\int d\lm_0\prod_{\al=1}^q d\lm_{\al}\xi_n^q(\lm_0) \lm_\al e^V
\prod_{\al=1}^q\eta_{\al,m_\al}(\lm_\al)=h_{n}Q_{\al,nm}
\prod_{\al=1}^q\delta_{m,m_\al}
\b
and the $P$-matrices as:
\a
\int d\lm_0\prod_{\al=1}^q d\lm_{\al}\xi_n^{q-1}(\lm_0)
{\d\over \d\lm_0}\left(\xi_n(\lm_0)e^{V_0+\lm_0\sum_{\al=1}^q} \right)
e^{\sum_{\al=1}^qV_\al}
\prod_{\al=1}^q\eta_{\al,m_\al}(\lm_\al)=\\
=P_{0,nm}h_{m}\prod_{\al=1}^q\delta_{m,m_\al}\0
\b

\a
\int d\lm_0\prod_{\al=1}^q  d\lm_{\al}\xi_n^q(\lm_0)
e^{V-V_\al-c_\al\lm_0\lm_\al}
\prod_{\al=1}^q
{\d\over \d\lm_\beta}\left(\eta_{\beta,m_\beta}(\lm_\beta)
e^{V_\al+c_\al\lm_0\lm_\al}\right)\prod_{\al\neq 0,\beta}
\eta_{\al,m_\al}(\lm_\al)=\0\\
=h_{n}P_{\beta,nm}\prod_{\al=1}^q\delta_{m,m_\al}\0
\b

We can now derive the coupling conditions:
\a
qP_0+V'_0(Q_0)+\sum_{\al=1}^qc_\al Q_\al&=&0\\
{\overline P_\al}+V'_\al(Q_\al)+c_\al Q_0&=&0,\ \ \al=1,\ldots q\0
\b
We consider only the symmetric case when the order of potentials is $p_\al=p$.
The calculation of degree for matrices gives:
\a
Q_\al &\in &[-m,n],\ \ \hbox{ for all }\ \al\0\\
Q_0 &\in &[-m_0,n_0],\
\b
where $m=1,n=p_0-1;m_0=p-1,n_0=1$.

The notation shows that we have $Q_\al$-matrices with finite band
with $m$ lower and $n$ higher diagonals.

We consider only 1-point correlation functions, hence we do not need
the flow equations. The 1-point correlation functions can be
calculated in the usual way as:
\a
<\Tr M_1^{k_1}\ldots M_q^{k_q}>=\Tr(Q_1^{k_1}\ldots Q_q^{k_q})
\label{cf}
\b
These 1-point correlation functions can be  calculated  in  every
genus $h$
and in principle for  arbitrary potentials, not only gaussian.

\section{The gaussian model}
\setcounter{equation}{0}
\setcounter{subsection}{0}
After the exact solution of $c=1$ matrix model (or chain q-multimatrix
model) with gaussian potential, the star  q-multimatrix  model  is  an
obvious target for study. Even with a trivial gaussian  potential,  it
could give an interesting string description. For example, due
to the additional permutation symmetry $S_q$, the tachyonic
field structure can be quite different from  the  original  chain
model.The 3-star matrix  model is also the
first one in the class of matrix models having the target space- the
$D_n$ Dynkin diagramm \cite{K}\cite{FK}.

The gaussian star-matrix model has the partition function:
\a
Z=\int \prod_{\al=1}^q dM_\al dM_0 \exp[t_0M_0^2+ u_0M_0+M_0
\sum_{\al=1}^q c_\al M_\al+\sum_{\al=1}^q(t_\al M_\al^2+
u_\al M_\al)]
\label{gp}
\b

The coupling conditions are:
\a
q P_0+2t_0Q_0+u_0+\sum_{\al=1}^{q}c_\al Q_\al &=&0\0\\
{\overline P_\al}+2t_\al Q_\al+u_\al+c_\al Q_0 &=&0,\ \ \al=1,\ldots q
\b

With the following parametrization of $Q$-matrices:
\a
Q_0&=&I_++a_0I_0+a_1\epsilon_-\\
Q_\al&=&b_\al/R_\al I_++d_\al I_0+R_\al\epsilon_-,\ \ \al=1,\ldots q\0
\b
we arrive at following equations:
\a
2t_\al R_\al+c_\al a_1=0\0\\
2t_\al b_\al+n+c_\al R_\al=0\0\\
2t_\al d_\al+u_\al+c_\al a_0=0\0\\
2t_0+\sum{c_\al b_\al\over R_\al}=0\\
2t_0a_0+u_0+\sum c_\al d_\al=0\0\\
2t_0a_1+qn+\sum c_\al R_\al=0\0
\b
Solving the coupling conditions we get :
\a
a_1&=&-{2q\over A},a_0={1\over A}(\sum{c_\al u_\al\over t_\al}-2u_0)\0\\
b_\al&=&-{1\over 2t_\al^2}({c_\al^2 q\over A}+t_\al),
R_\al={c_\al q\over t_\al A}\0\\
d_\al&=&{1\over At_\al}(c_\al u_0-2t_0u_\al+u_\al\sum{c_\beta^2\over 2t_\beta}-
c_\al\sum{c_\beta u_\beta\over 2t_\beta})\0
\b
where $A=4t_0-\sum c_\al^2 / t_\al$.

For quadratic potentials $V_\al$, the basic functions are
 Hermite polynomials:

\a
\xi_n(\lm_{0})&=&\eta_n(\lm_0)= H_n({\lm_{0}-a_0\over \sqrt{2a_1}}),\\
\eta_{\al,m}(\lm_{\al})&=& H_n({\lm_{\al}-d_\al\over
\sqrt{2b_\al}}),\al=1\ldots q\0
\b
To calculate the partition function we observe that :
\a
h_{0,n}={1\over A},h_{\al,n}=R_{\al}^n
\b
giving:
\a
Z=const\left({1\over A}\prod_{\al=1}^q {c_\al\over
t_\al}\right)^{{N^2\over 2}}
\b
 $const$ in our case is the exponent
$\exp[-1/4(Aa_0^2+\sum_\al u_\al^2/t_\al)]$ obtained by shifting the
 matrices $M_\al$ so that the liniar terms in the potential vanish.

The partition function and the first simplest 1-point correlation
functions as $\Tr Q_\al^2,\Tr Q_\al$ can be calculated by  direct
integration of the original integral (\ref{gp}). This is not  the
case for more complicated 1-point correlation
functions  as  those  given  by   (\ref{cf}).Instead   with   the
$Q$-matrix approach this is  easy , using  the  explicit  form  of
$Q$-matrices.In the Dyson-Schwinger approach ,this calculation
is also possible but only on the sphere.

As an example we give the result for the 1-point correlation function
(when $u_0=u_\al=0$ and $a_0=d_\al=0$):
\a
&&\Tr Q_\al^nQ_\beta^m=\sum_{2k=0}^n\sum_{2l=0}^m\sum_{i=0}^{n-2k}
\sum_{j=0}^{m-2l}
{n!m!(-1)^k2^{-(l+k)}\over i!j!k!l!(n-2k-i)!(m-2l-j)!}\\
&&b_\al^{n-(i+k)}b_\beta^{m-(j+l)}R_\al^{2(i+k)-n}R_\beta^{2(j+l)-m}
\left(i+j\right)!\left(\ba{c} N+i+j\\i+j+1\ea\right)
\delta({n+m\over 2}-(l+k),i+j)\0
\b
In the large $N$ limit we have $l+k=0$ or $l=0,k=0,i+j={n+m\over 2}$ and
the previous formula simplifies to:
\a
\Tr Q_\al^nQ_\beta^m=\sum_{i=0}^{n}\sum_{j=0}^{m}
{n!m!\over i!j!(n-i)!(m-j)!}
b_\al^{n-i}b_\beta^{m-j}R_\al^{2i-n}R_\beta^{2j-m}
{N^{{n+m\over 2}+1}\over{n+m\over 2}+1}
\b
For $m=0$ and $n=2r$ we obtain the correlation function in genus 0:
\a
\Tr Q_\al^{2r}={(2r)!\over r!(r+1)!}N^{r+1}(b_\al )^r
\b
We have sum over $i=r,j=0$.

\section{The gaussian model on Bethe tree}
\setcounter{equation}{0}
\setcounter{subsection}{0}

Kazakov and Migdal \cite{KM} have obtained the so-called induced "induced
QCD"-
a matrix model embedded in the  regular  D-dimensional  lattice.  For
gaussian potential , the model was  solved  by  Gross \cite{Gross}.
 In  the  limit
$N\rightarrow \infty$ the Kazakov-Migdal model with generic  potential
is equivalent to the matrix model  with  a  Bethe  tree  target  space
\cite{Boulatov}.

 Because  the  model  was  studied  only   with   the
saddle-point method, it is  interesting  to  study  it  in  a different
framework , that of Q-matrices approach. It gives higher  accuracy  in
studying different regions of Bethe tree and also permits computations
in higher genera.

Our  model is the inhomogenous version of matrix model on Bethe  tree,
at every level of branch we assigne a specific partition function  and
propagator.

The gaussian matrix model on Bethe tree is ($i=2j$):
\a
Z=\int dM_i \exp \Tr[\sum_i (t_i M_i^2+ u_i M_i)+
\sum_{<ij>}c_i M_i M_{j}]
\b
where $<ij>$ denotes the permitted links of Bethe tree, and $c_i$  are
the coupling constants of the  i-th level branch.

The coupling conditions are:
\a
q P_i+2t_iQ_i+u_i+c_iQ_{i-1}+(q-1)c_{i+1}Q_{i+1} &=&0\0\\
q{\overline P_{i+1}}+2t_{i+1} Q_{i+1}+u_{i+1}+c_{i+1} Q_i +(q-1)
c_{i+2}Q_{i+2}&=&0
\b
Introducing $Q$-matrices with the form:
\a
Q_i&=&S_i I_++a_i I_0+f_i\epsilon_-\\
Q_{i+1}&=&b_i/R_i I_++d_i I_0+R_i \epsilon_-,\0
\b
 we get from the coupling conditions the following equations for coefficients:
\a
2t_i S_i+c_i{b_{i-1}\over R_{i-1}}+(q-1)c_{i+1}{b_i\over R_i}=0\0\\
2t_ia_i+u_i+c_id_{i-1}+(q-1)c_{i+1}d_i=0\0\\
2t_if_i+{qn\over S_i}+c_iR_{i-1}+(q-1)c_{i+1}R_i=0\\
2t_{i+1}R_i+c_{i+1}f_i+(q-1)c_{i+2}f_{i+1}=0\0\\
2t_{i+1}d_i+u_{i+1}+c_{i+1}a_i+(q-1)c_{i+2}a_{i+1}=0\0\\
2t_{i+1}{b_i\over R_i}+{qn\over R_i}+c_{i+1}S_{i}+(q-1)c_{i+2}S_{i+1}=0\0\
\b
Let suppose that all coefficients for various $i$ are equal
$c_i=c,t_i=t,u_i=u$.In this case
the equations reduce to the following set of equations:
\a
f(S_i)+g(R_{i-1})=0,\qquad f(R_i)-g(S_{i})=0\0\\
f(a_i)+K=0,\qquad f(d_i)+K=0\\
f({b_i\over R_i})-{2qnt\over c^2}{1\over R_i}=0,
\qquad f(f_i)-{2qnt\over c^2}{1\over S_i}=0\0
\label{fgeq}
\b
where the constant $K=(q-{2t\over c})u$ and the functions $f,g$ are:
\a
f(x_i)&=&x_{i-1}+\left[2(q-1)-4\left({t\over c}\right)^2\right]x_i
+(q-1)^2x_{i+1}\0\\
g(x_i)&=&{qn\over c}({1\over x_i}+{q-1\over x_{i+1}})
\b

\subsection{Fractal regime}

 In the scaling limit $R_i\rightarrow R,
S_i\rightarrow S, b_i\rightarrow b, f_i\rightarrow f, d_i\rightarrow
d, a_i\rightarrow a$ and (we take $n=1$):
\a
RS={q^2c\over 4t^2-(qc)^2},\0\\
b=fS=-{qt\over 4t^2-(qc)^2}\\
a=d={qc-2t\over 4t^2-(qc)^2}u\0
\b

This case represents the limiting case of a matrix model on the fractal curve
of Bethe tree. We can
define the free energy on the unit length:
\a
F_{frac}=\log RS= \log {q^2 c \over 4t^2-(qc)^2}
\b
We observe a singularity of the free energy at $(2t/c)^2=1$. This critical
point is the analog of critical point for $c=1$ matrix model (or
$q$-multimatrix
model) at the self-dual radius $R^2=1$
We must consider the physical domain when $2t<qc$ ($t<0$ to have a well-defined
path-integral).
We choose $c<0$. We see that the other region $2t>qc, c>0$ is not reached.

We can calculate the 1-point correlation function
 (for simplicity we take $u=0$ and genus $0$)
        \a\left<\Tr M_i^{2k}\right>_0=\left<\Tr M_{i+1}^{2k}\right>_0
        &=& \Tr(Q^{2k})
                = b^k\Tr(I_++\epsilon_-)^{2k}\0\\
                &=&{(2k)! N^{k+1}\over k!(k+1)!}
                  \left({-qt\over 4t^2-(qc)^2}\right)^k\b

\subsection{Asymptotic regime}

We have considered the fractal curve or a tiny strip of surface which is
filling densily
the extremity of the Bethe tree. We can study a larger strip which tends
asymptotically
to the fractal curve.

In the large $N$ limit we can scale the coefficients
$R_i\rightarrow \sqrt{N}R_i(x), S_i\rightarrow \sqrt{N}S_i(x),
b_i\rightarrow Nb_i(x), f_i\rightarrow \sqrt{N}f_i(x),
 d_i\rightarrow d_i(x), a_i\rightarrow a_i(x)$, where $x=n/N$.
We can see now that all second terms in the equations (\ref{fgeq})
 are proportional with $x, 0\leq x<1$ and can be considered as
perturbations. We neglect the function $g$.
In this case we can solve the recursion relations
(\ref{fgeq}) with the result:
\a
R_i&=&r^i,S_i=s^i,\0\\
a_i&=&r^i-{2t\over q^2c}K,d_i=s^i-{2t\over q^2c}K,\ K={qcu\over 2t}-u\\
b_i&=&b^i r^i,f_i=f^i\0
\label{power}
\b
where $r=r_+,s=r_-$ or viceversa with $r_\pm$ being the solution of second
order eq.:
\a
(q-1)^2 r_\pm^2+(2(q-1)-{4t^2\over c^2})r_\pm+1=0
\label{sec}
\b
and $b=b_+,f=b_-$ or viceversa with $b_\pm$ being the solution of second order
eq.:
\a
(q-1)^2 b_\pm^2+(2(q-1)+{2tq x\over c^2}-{4t^2\over c^2})b_\pm+1=0
\b

We see that the alternating $Q_i, Q_{i+1}$ in the asymptotic regime
can be interchanged;
hence it does not matter if $i$ is odd or even.

We can define also in this regime the free energy on the unit length:
\a
F_{i,i+2}=\log (q-1)^{2i}R_iS_i= 0
\b
(q must be bigger than 2 to have at least one branch).The free energy
is 0 and is different from $F_{frac}$. This is understandable because
$F_{frac}$ is proportional with $x$, but we have considered the case when
$x=0$, hence $F_{i,i+2}=0$. To see if $F_{i,i+2}$ really tends to
$F_{frac}$ we must include the perturbation in $x$.

As we said before, our  model  differs  from  the  one  studied by  Gross  and
Boulatov \cite{Gross}\cite{Boulatov}. Their model is  homogenous  ;for
the Bethe lattice with coordonation number $2D$ their partition function
is:
\a
Z=\int dM_i\exp\Tr[-\sum_i {m^2\over 2}M_i^2+\sum_{<ij>}M_iM_j]=
\int dM Z(M)^{2D}\exp\left(-{m^2\over 2}\Tr M^2 \right)\0
\b
where  the partition  function  of  a  branch  $Z(M)$  satisfies  the
equation:
\a
Z(M)=\int dM' Z(M')^{2D-1}\exp\Tr(-{m^2\over 2}M'^2+M'M)\0
\b
Our model is inhomogenous. Every branch of different level $i$  has  a
different partition  function  $Z_i(M)$.  Hence  the  total  partition
function is:
\a
Z=\int dM Z(M)^{q}\exp\left(-{t\over c}\Tr M^2\right)\0
\b
and $Z_i(M)$ satisfies the equation:
\a
Z_i(M)=\int dM' Z_{i+1}(M')^{q-1}Z_{i-1}(M')\exp\Tr(-{t\over c}M'^2+M'M)
\label{treeq}
\b
We see that the partition function for $i$-th level branch is  expressed
not only in terms of higher level  branches ($i+1$),  but  we  have  also  the
back-reaction on the lower level branches ($i-1$).We also observe that
the eq.( \ref{treeq}) is different for $i$ odd or  even.

 Solving  the equation is equivalent with the first two equations
(\ref{fgeq}). Solving them as recursion equations we get the power-like
 solution (\ref{power})
$Z_j(M)=r^{j\Tr M^2}  $, where $r$ satisfies the second order equation
(\ref{sec}). If we solve (\ref{fgeq}) as differential equations (after
proper scaling when $|(i+1)-i|\ll i$) we get the exponential solution
$Z_j(M)=\exp (j r \Tr M^2) $ where $r$ is (from equation (\ref{sec})):
\a
r_\pm={2\left({t\over c}\right)^2-(q-1)\pm 2\left({t\over c}\right)
 \sqrt{\left({t\over c}\right)^2
-(q-1)}\over (q-1)^2}\0
\b
The signs alternate for $i$ odd and even.
This result can be compared  with  that  of  homogenous  model  if  we
identify $t/c=m^2/2,q=2D$.  The  partition  function  per  branch  for
homogenous model behaves as $Z(M)=\exp (-\al\Tr M^2)$, where $\al$ is:
\a
\al_\pm={m^2(D-1)\pm D\sqrt{m^4-4(2D-1)}\over 2D-1}\0
\b
We have an interesting  property  in  the  asymptotic  regime:the
coefficient $R_k$ near the point ${1\over q-1}(t/c)^2=1/2$ has  a
slow oscilation with the period $T_{\Delta k}$.

The coefficient  $R_k$ is directly related with the free energy of
branches of level between $k$ and $k+1$:
\a
F_{k,k+1}=\log (q-1)^{2k}R_k\0
\b
$S_k$ has a complementary oscilation such that
$F_{k,k+1}(R_k)+F_{k+1,k+2}(S_k)=F_{k,k+2}=0$.
This behaviour is typical to an antiferromagnet: $R_k, S_k$ are like
spin-up and spin-down configurations which group pair-wise having a
total energy zero.

We take for convenience $R_0=1,R_1=0$. The coefficient $R_k$ is:
\a
R_k={r_+^{k+1}-r_-^{k+1}\over r_+-r_-}\0
\b
where $r_\pm$ are the solutions of the equation (\ref{sec}).
We introduce the notation $\beta={1\over q-1}(t/c)^2$ and consider the
region $0\leq \beta\leq 1$. In this case:
\a
r_\pm={1\over (q-1)^2}e^{\pm \omega},\ \ \
\arctan\omega={2\sqrt{\beta(1-\beta)}\over 2\beta-1}
\label{r}
\b
In this region $R_k=(q-1)^{-2k}{\sinh k\omega \over \sinh\omega}$ has a fast
decaying in amplitude (for $q>2$) and an oscilatory character with
pulsation $\omega$ . For $\beta\sim {1\over 2}$ the expression (\ref{r}) is
singular
and the pulsation is $\omega\sim {\pi \over 2}$. If we take $\beta={1\over
2}+\epsilon$ then $\omega={\pi \over 2}+\Delta\omega$ and
$\Delta\omega\sim\epsilon$. This property induces a modulation of the
oscilation
with the period :
\a
T_{\Delta k}\sim\left(1-{2\over q-1}\left({t\over c}\right)^2\right)^{-1}
\b
For $\beta\rightarrow 1/2$ the modulation dissapears because the period
$T_{\Delta k}\rightarrow \infty$.

\section{The q-Potts-like model}
\setcounter{equation}{0}
\setcounter{subsection}{0}

q-state Potts spins are an interesting generalization  of  the  Ising
model ($q=2$) . On planar random lattice they were studied first time by
Kazakov \cite{Kaz} \cite{KM1}. The  cases  $q\rightarrow  0$  and
$q\rightarrow 1 $ represent the models of tree-polymers,  respectively
that of percolation. The $q=2$ case , that of Ising  model  on  random
lattice, can be expressed in terms of the 2-matrix model.

The partition function of q-state Potts model on a random lattice is:
\a
Z(g,\beta,H)=\sum_n g^n \sum_{\{G^{(n)}\} }\sum_{\{\sigma\} }
\exp[-{\beta\over 2} \sum_{k,j}^n G_{kj}^{(n)}(\delta_{\sigma_k\sigma_j}-1)
+H\sum_k^n(\delta_{1,\sigma_k}-1)]
\label{pottsspin}
\b
where the summ run over all triangulations with $n$ triangles $\{G^{(n)}\}$
and all  spin  configurations  $\{\sigma\}$.  $G_{kj}^{(n)}$-  is  the
adjancency matrix  of  planar  lattice  with  $n$  vertices,  $\beta$-
inverse temperature, $H$- magnetic field .

For  zero  magnetic  field,
the partition function can be expressed in terms of the matrix model:
\a
Z=\int \prod_{\al=1}^q dM_\al\exp[2c\sum_{\al>\beta}^q
M_\al M_\beta -\sum_{\al=1}^q(M_\al^2+g/3 M_\al^3)]
\label{pottsmat}
\b
The partition functions  (\ref{pottsmat})  and  (\ref{pottsspin})  are
equal due to the equivalence of the Feynman graphs of matrix model  on
dual   lattice    with    inverse    temperature    $\beta_{dual}=\log
(1+q(e^\beta-1)^{-1})$ and the Boltzmann weights of Potts model on  the
original lattice.

Introducing a new matrix $M_0$ the previous integral  can  be  rewritten
as:
\a
Z=\int \prod_{\al=1}^q dM_\al dM_0 \exp[\tilde{c}M_0\sum_{\al=1}^q
M_\al -M_0^2/2-\sum_{\al=1}^q(M_\al^2/2-\tilde{g}/3 M_\al^3)]
\b
with  $\tilde{c}^2=c/(1+c),\tilde{g}^2=g^2/(2(1+c))^3$.  The  coupling
constant $c$ is connected with the inverse temperature $\beta$ by  the
formula
\a
c^2=(e^\beta+q-1)^{-1}\0
\label{temp}
\b

We consider the star-matrix model with partition function:
\a
Z=\int \prod_{\al=1}^q dM_\al dM_0 \exp[t_0M_0^2+ u_0M_0+M_0
\sum_{\al=1}^q c_\al M_\al+\0\\
+\sum_{\al=1}^q(s_\al M_\al^3+t_\al M_\al^2+
u_\al M_\al)]
\label{partmat}
\b
The coupling conditions are:
\a
q P_0+2t_0Q_0+u_0+\sum_{\al=1}^{q}c_\al Q_\al &=&0\0\\
{\overline P_\al}+3s_\al Q_\al^2+2t_\al Q_\al+u_\al+c_\al
Q_0 &=&0,\al=1,\ldots q\0
\b
With the following parametrization of $Q$-matrices:
\a
Q_0&=&I_++a_0I_0+a_1 I_-+a_2 I_{-2}\0\\
Q_\al&=&b_\al/R_\al I_++d_\al I_0+R_\al I_-,\al=1,\ldots q\0
\b
we arrive at following equations:
\a
\left.\ba{r} 3s_\al R_\al^2+c_\al a_2=0\\
6s_\al R_\al d_\al+2t_\al R_\al+c_\al a_1=0\\
6s_\al b_\al d_\al+2t_\al b_\al+n+c_\al R_\al=0\\
3s_\al (d_\al^2+2b_\al)+2t_\al d_\al+u_\al+c_\al a_0=0
\ea \right\}, \ \al=1,\ldots q
\label{copl1}
\b
\a
\left.\ba{r}
2t_0+\sum{c_\al b_\al\over R_\al}=0\\
2t_0a_0+u_0+\sum c_\al d_\al=0\\
2t_0a_1+qn+\sum c_\al R_\al=0
\ea \right\}
\label{copl2}
\b
\subsection{Symmetric case}

We solve the special  symmetric  case  when  $s_\al=s,t_\al=t,u_\al=u,
c_\al=c$. In this case
$R_\al=R,d_\al=d,b_\al=b$ .

We can express all the coefficients in terms of only two of  them  $R$
and $d$:

\a
a_2=-{3s\over c}R^2, a_1=-{q(n+cR) \over 2t_0}\0\\
a_0=-{u_0+qcd\over 2t_0},b=-{2t_0R\over qc}
\b
 $R$ and $d$ satisfy a system of 2 non-linear equations:
\a
d&=&-{t\over 3s}+{cq\over 12s t_0}(c+{n\over R})\0\\
R&=&{qc\over 4t_0}d^2+({qct\over 6st_0}-{(qc)^2c \over 24st_0^2})d+{qcu\over
12st_0}
-{(qc)cu_0\over 24st_0^2}
\b
Hence all coeficients of the $Q$-matrices can be expressed in terms of the
$R_\al=R$
coeficient which satisfies a third-order equation:
\a
R^3+R^2 {qc\over (12st_0)^2}\left({(4tt_0-qc^2)^2\over
4t_0}+6s(cu_0-2t_0u)\right)-{(qc)^3n^2\over 4t_0(12st_0)^2}=0
\label{rvar}
\b
For symplicity we choose $6s=2t_0=2t=-1,u_0=0,c^2=1/q$. If we denote
$z=R/qc$ we have instead (\ref{rvar}) the equation:
\a
z^3-z^2u+1/2=0
\label{z}
\b
Two roots coincide $z_1=z_2=z_*=1$ at  the  critical  point  when  the
"cosmological constant" $u_*=3/2$. Near critical point,  the  variable
$R$ related with the free energy will scale as
 $R-R_*\sim (u-u_*)^{2/(p+q-1)}$ for the $(p,q)$ matter models coupled
with the 2d gravity.

Expanding $u$ and $z$ near critical point:
\a
u=u_*+\mu\delta^2,z=z_*+Z\delta^2\0
\b
we get in the lowest order of $\delta$ the relation  $\mu=3Z^2$.  This
means that the variable $R$ scales as  $R-R_*\sim (u-u_*)^{1/2}$.This
critical point (the continuum limit) corresponds to the pure gravity model
or $\phi^3$ 1-matrix model.The results remain true for arbitrary
$s,t_0,t,u_0,c$ in the symmetric case.

\subsection{Non-symmetric case}

We can write the system of equations (\ref{copl1}), (\ref{copl2})
 in a different way
which will permit to remain with a single type of variables $X_\al$:
\a
X_\al=3s_\al d_\al+t_\al'
\b
We denote by:
\a
u_\al'=u_\al-{c_\al u_0\over 2t_0},c_{\al\beta}=-{c_\al c_\beta\over 4t_0},
t_\al'=t_\al-{c_\al^2\over 4t_0}\0
\b
Then we can express the variables $R_\al$ in terms
of $X_\al$ from the system:
\a
X_\al R_\al+\sum_{\beta \neq \al}c_{\al\beta}R_\beta={qn\over 4t_0}
,\ \al=1\ldots q
\label{R}
\b
and also the variables $b_\al$ in terms of $X_\al$:
\a
{2b_\al X_\al+n\over R_\al}+\sum_{\beta \neq \al}c_{\al\beta}
{2b_\beta \over R_\beta}=0,\ \al=1\ldots q
\label{b}
\b
We remain with the system:
\a
{X_\al^2\over 3s_\al}+3s_\al(2b_\al)+u_\al'-{t_\al'^2 \over 3s_\al}+
\sum_{\beta \neq \al}{2c_{\al\beta}\over 3s_\al}(X_\beta-t_\beta')=0
,\ \al=1\ldots q
\label{xx}
\b
We have the suplimentary constraint which can be imposed on
the variables $X_\al$:
\a
{3s_\al\over c_\al}R_\al^2=const,\ \al=1\ldots q
\label{constr}
\b
When we tend to the symmetric case with $c_{\al\beta}=c,t_\al'=0,
6s_\al=2t_0=-1,u_\al'=u+c^2(q-1)^2$ we get from the system
(\ref{constr}) the following equation:
\a
2(X+c(q-1))^2-{1\over 2(X+c(q-1))}-u=0\0
\b
With the notation $z=R/q=-1/(2X+2c(q-1))$ we obtain the equation
(\ref{z}).

{\bf q=2 case}

We  argue that  the critical behaviour of the  case  $q=2$
 coincides with   the Ising model on random $\phi^3$ lattice.

Solving the system (\ref{b}) we get:
\a
2b_1=-{n(X_2-cr)\over X_1X_2-c^2},\
2b_2=-{n(X_1-c/r)\over X_1X_2-c^2}\0
\b
where:
\a
r={R_1\over R_2}=\left({c_1 s_2\over c_2 s_1}\right)^{1/2}
\label{r}
\b
Because $R_1,R_2$ are:
\a
R_1=-{(X_2-c)\over X_1X_2-c^2}{2n\over 4t_0},\
R_2=-{(X_1-c)\over X_1X_2-c^2}{2n\over 4t_0}\0
\b
the relation (\ref{r}) can be rewritten as:
\a
r={X_2-c\over X_1-c}\0
\b
We remain with the system:
\a
{X_1^2\over 3s_1}+{2c\over 3s_2}X_2-3s_1n{X_2-cr\over
X_1X_2-c^2}+u_1' &=&{t_1'^2\over 3s_1}+2c{t_2'\over 3s_2},\\
{X_2^2\over 3s_2}+{2c\over 3s_2}X_1-3s_2n{X_1-c/r\over X_1X_2-c^2}+u_2'
&=&{t_2'^2\over 3s_2}+2c{t_1'\over 3s_1}\0
\label{Ising1}
\b
We point out the great similarity between this system and that of
the Ising model on $\phi^3$ lattice \cite{potts} (which corresponds to
 the case $q=2$ Potts model).
To show this, we integrate the intermediate matrix $M_0$ in the  relation
(\ref{partmat}).We get the two-matrix model:
\a
Z=\int dM_1dM_2\exp[\sum_{\al=1}^2(s_\al M_\al^3+t_\al' M_\al^2+u_\al'
M_\al)+2c M_1 M_2]\0
\b
with the previous notations for $t_\al',u_\al',c=c_{12}$.

Solving the coupling conditions we remain with three equations:
\a
X_1X_2 &=& c^2+{nc\over 2R},\0\\
{X_1^2\over 3s_1}+({2c\over 3s_2}-{3s_14R\over 2c})X_2+u_1' &=&
{t_1'^2\over 3s_1}+2c{t_2'\over 3s_2},\\
{X_2^2\over 3s_2}+({2c\over 3s_2}-{3s_24R\over 2c})X_1+u_2' &=&
{t_2'^2\over 3s_2}+2c{t_1'\over 3s_1}\0
\label{Ising2}
\b
Introducing the expression of $2R=nc/ (X_1X_2-c^2)$  in the  last
two equations ,we see that they differ from the equations (\ref{Ising1}) only
by the terms containing the $r$ quantity. We expect that  these  terms
are only an artefact of the different basis of orthogonal polynomials
we have  chosen  and that they do not change the critical behaviour of
the free energy. The variable $R$ scales at the critical point as
$R-R_*\sim (u-u_*)^{1/3}$ (we take $u_1'=u,u_2'=0$).

{\bf q=3 case}

This is the first non-trivial member of the q-Potts-type set  of
models.It can not be derived from the two-matrix model or other more
complicated chain models.It represents the matrix model on the Dynkin
diagramm $D_3$.

We have not managed to derive the critical behaviour.However, for
further developments, we write
down the system of equations which gives the critical scaling.

We introduce the function:
\a
Y_{123}(1,r_1,r_2)=X_2X_3-c_{23}^2+r_1(c_{13}c_{23}-c_{12}X_3)+
r_2(c_{12}c_{23}-c_{13}X_2)
\b
The indices $123$ of the function $Y_{123}$ are related with the indices of
the variables $X_\al,c_{\al\beta}$.
 The function is symmetric only in the last two indices:
$Y_{123}(1,r_1,r_2)=Y_{132}(1,r_2,r_1)$.

We also introduce:
\a
Y=X_1X_2X_3+2c_{12}c_{13}c_{23}-(c_{13}^2X_2+c_{23}^2X_1+c_{12}^2X_3)\0
\b

Then the coupling conditions (\ref{xx}) in this case are (for
simplicity we consider $t_\al'=0$):
\a
{X_1^2\over 3s_1}+{2c_{12}\over 3s_2}X_2+{2c_{13}\over 3s_3}X_3-3s_1n\;\;\;
{Y_{123}(1,r_1,r_2)\over Y}\;\;\;\;\;\;+u_1' &=&0
,\0\\
{X_2^2\over 3s_2}+{2c_{12}\over 3s_1}X_1+{2c_{23}\over 3s_3}X_3-3s_1n
{Y_{231}(1,r_2/r_1,1/r_1)\over Y}+u_2' &=&0
,\\
{X_3^2\over 3s_3}+{2c_{13}\over 3s_3}X_1+{2c_{23}\over 3s_2}X_2-3s_1n
{Y_{312}(1,r_1/r_2,1/r_2)\over Y}+u_3' &=&0
,\0
\b
The variables $r_1,r_2$ can be written in terms of $X_\al$:
\a
r_1={Y_{123}(1,1,1)\over Y_{231}(1,1,1)},
r_2={Y_{123}(1,1,1)\over Y_{312}(1,1,1)}
\b

\section{Conclusions}
\setcounter{equation}{0}
\setcounter{subsection}{0}
We have studied the inhomogenous matrix model on the Bethe tree and
have obtained similar results with the homogenous model in the
saddle-point method.We have two regimes: the fractal and the
asymptotic .In the asymptotic regime, we get the exponential behaviour for the
partial
partition function $Z_j(M)=\exp(jr\Tr M^2)$, where $r$ satisfies a
second order equation like in the homogenous case. For large $j$- the
level of the branch- we expect that the properties of the model become
independent of $j$ and is a transition from the asymptotic regime to
the fractal regime.
Also in the asymptotic regime when $\beta={1\over q-1)}(t/c)^2=1/2$ we have
a slow oscilation of the free energy with the period $T\sim
(1/2-\beta)^{-1}$.

For the q-Potts-like model with arbitrary $q$ we write down the
general coupling conditions. For the special cases $q=2$ and $q=3$ we
solve the coupling conditions in terms of only one type of variables
$X_\al$. For  $q=2$ we have argued that the system has the critical
behaviour of the Ising model on $\phi^3$ lattice .

\begin{center}
AKNOWLEDGEMENTS
\end{center}
I  would  like  to  thank  Prof.L.Bonora  and Dr.F.Nesti  for  many  usefull
discussions. I also want to aknowledge the SISSA  staff  and  students
for the stimulating enviroment.

\end{document}